\documentclass[aps,prc,twocolumn,superscriptaddress,showpacs,floatfix,nofootinbib,preprintnumbers]{revtex4-1}
\usepackage{amsmath,graphicx,color,hyperref}

\newcommand{\trento}{T$\mathrel{\protect\raisebox{-2.1pt}{R}}$ENTo}

\begin{document}

\title{The size of the quark-gluon plasma in ultracentral collisions: impact of initial density fluctuations on the average transverse momentum}

\author{Fabian Zhou}
\affiliation{Institute for Theoretical Physics, University of Heidelberg, 69120 Heidelberg, Germany}
\author{Giuliano Giacalone}
\affiliation{Theoretical Physics Department, CERN, CH-1211 Gen\`eve 23, Switzerland}
\author{Jean-Yves Ollitrault}
\affiliation{Institut de physique th\'eorique, Universit\'e Paris Saclay, CNRS, CEA, F-91191 Gif-sur-Yvette, France} 
\date{\today}

\begin{abstract}
Recent experiments have shown that the mean transverse momentum $\langle p_T\rangle$ of outgoing particles increases as a function of the particle multiplicity in ultracentral nucleus-nucleus collisions at collider energies. This increase was originally predicted on the basis of simulations where the multiplicity increase occurred at constant volume, so that it implied a larger density and temperature. However, recent state-of-the-art simulations have shown that, for some models of initial condition, the volume may vary with the multiplicity in ultracentral collisions. We elucidate this effect by analytically relating the variation of the volume  to the radial distribution of the one- and two-point functions of the fluctuating density field. We show that the volume variation is small if the total entropy of the ultracentral collisions scales with the mass number of the colliding isotopes. We argue that probing detailed transverse distributions of initial-state fluctuations through the ultracentral $\langle p_T\rangle$ has nontrivial implications for models of nuclear structure and of the pre-equilibrium stages.
\end{abstract}

\preprint{CERN-TH-2025-228}

\maketitle

\section{Introduction}

Ultracentral Pb+Pb collisions at the Large Hadron Collider (LHC), typically defined as the 0.2\% fraction producing the largest number of particles~\cite{CMS:2013bza}, open a unique window on the initial stages of the collision~\cite{Luzum:2012wu,Shen:2015qta}. 
The reason is that they have essentially zero impact parameter~\cite{Das:2017ned}, so that two different ultracentral collisions differ only by quantum fluctuations~\cite{Samanta:2023amp}. 
These quantum fluctuations originate in particular from the wavefunctions of colliding nuclei~\cite{Liu:2022kvz,Zhang:2025voj}.
They are characterized, to leading order, by the two-point function of the initial entropy density field~\cite{Blaizot:2014nia}. 
This two-point function can be related to two-body correlations within the nucleus~\cite{Giacalone:2023hwk,Mehrabpour:2025ogw}, and this relation paves the way to detailed theoretical predictions~\cite{Duguet:2025hwi,Liu:2025uks}. 
But there are at present few constraints from heavy-ion experiments, if any, on how fluctuations are distributed through the transverse plane. 
The only existing constraint is on the variance of the total entropy at zero impact parameter, which is inferred from the tail of the multiplicity distribution~\cite{Yousefnia:2021cup,Pepin:2022jsd}, and corresponds to the integral of the two-point function. 

We show that the increase of the mean transverse momentum $\langle p_T\rangle$ as a function of the collision multiplicity $N_{ch}$, which has recently been observed in ultracentral Pb+Pb collisions at the LHC~\cite{CMS:2024sgx,ATLAS:2024jvf,ALICE:2025rtg}, can be used to constrain the spatial dependence of the two-point function. 
Hydrodynamic simulations have shown that $d\ln\langle p_T\rangle/d\ln N_{ch}=c_s^2$, where $c_s^2$ is the sound velocity at an effective temperature which will be defined in Sec.~\ref{s:trajectum}. 
This relation is surprisingly accurate for a broad range of collision energies (encompassing that of the LHC), freeze-out temperatures~\cite{Gavassino:2025bts} and equations of state resembling that of QCD~\cite{Gardim:2024zvi}, provided that the transverse size of the quark-gluon plasma produced in the early stages of the collision is independent of $N_{ch}$. 
The physical picture is that an increase in $N_{ch}$ corresponds to an increase in entropy density, while the increase of $\langle p_T\rangle$  measures that of the corresponding temperature~\cite{Gardim:2019brr}, and the two are related by the compressibility, which defines the speed of sound~\cite{Ollitrault:2007du}. 
But the hypothesis of constant size has been criticized on the basis of state-of-the-art simulations~\cite{Nijs:2023bzv}, which show that depending on the details of the initial-state model, the quark-gluon plasma may shrink or swell as the multiplicity increases~\cite{Sun:2024zsy}. 

We study the size of the quark-gluon plasma in ultracentral collisions and its dependence on the model of initial conditions. 
We start by recalling in Sec.~\ref{s:basics} the motivations behind the default model of initial conditions, which postulates that the  entropy density at the beginning of the hydrodynamic evolution is determined, at a given point of the transverse plane, by $s\propto (t_At_B)^{0.5}$~\cite{Moreland:2014oya},  where $t_A$ and $t_B$ are the thickness functions (nuclear density integrated over longitudinal coordinate) of colliding nuclei at that point. 
We then describe (Sec.~\ref{s:trento}) Monte Carlo simulations with exponents $\nu$ that vary around $0.5$~\cite{Nijs:2023yab,Nijs:2023bzv}. 
In Sec.~\ref{s:mc}, we show that the size of the quark-gluon plasma can increase or decrease as a function of the multiplicity depending on the value of $\nu$, while   it remains constant for the default model ($\nu=0.5$). 
In Sec.~\ref{s:perturbative}, we explain these results by explicating the relation between the system size and the two-point function of the initial entropy density.
The default model is worked out analytically in appendices, where we prove that it leads to a constant size (App.~\ref{s:analytic}) and discuss the effect of the nucleon width on the mean density profile (App.~\ref{s:wp}). 
In Sec.~\ref{s:trajectum}, we finally argue that the increase of $\langle p_T\rangle$ in ultracentral collisions can be used to infer information about local density fluctuations. 
 
\section{Understanding $s\propto\sqrt{t_At_B}$}
\label{s:basics}

The crucial quantity for phenomenology is the entropy density at the time $\tau_h$ when the system thermalizes. 
It serves as an initial condition for hydrodynamic calculations and determines multiplicities and spectra of outgoing particles.  
We assume for simplicity that it is invariant under longitudinal boosts~\cite{Bjorken:1982qr}, and we denote it by $s({\bf x})$, where ${\bf x}$ is the transverse coordinate. 
We denote by $S$ its integral over the transverse plane, which we loosely refer to as the total entropy:\footnote{Strictly speaking, $\tau_h S$ is the entropy per unit rapidity.}
\begin{equation}
  \label{defS}
  S\equiv \int_{\bf x} s({\bf x}).
\end{equation}
In phenomenological studies, the total entropy is typically adjusted to match the observed multiplicity, and it not part of the model itself. 
The modeling lies in the variation of the entropy density with ${\bf x}$ and, more specifically, how it depends on the thickness functions $t_A({\bf x})$ and $t_B({\bf x})$ (which will be defined more precisely in Sec.~\ref{s:trento}) of incoming nuclei. 

In this Section, we briefly recall why the scenario $s\propto\sqrt{t_At_B}$ is preferred, both from the point of view of comparison with experimental data, and from general theoretical considerations. 

It has long been known that the multiplicity produced in a heavy-ion collision is proportional to the number of nucleons (more precisely, to the number of quarks) involved in the collision~\cite{Eremin:2003qn,STAR:2015mki}. 
Models of the initial entropy density have been elaborated which take into account this constraint by imposing that $s$ is homogeneous of degree 1 in $t_A$ and $t_B$, e.g., $s\propto (t_A^p+t_B^p)^{1/p}$~\cite{Moreland:2014oya}. 
Theory to data comparison systematically favors values of $p$ close to $0$ for the entropy density \cite{Bernhard:2016tnd,Nijs:2023yab}, corresponding to $s\propto\sqrt{t_At_B}$. 
In particular, this value reproduces well the centrality dependence of elliptic flow and of the measured particle yields.\footnote{For this reason, the same value is also favored when the \trento{} model is used to initialize the energy density at midrapidity~\cite{Nijs:2020ors,Nijs:2020roc,JETSCAPE:2020mzn,Parkkila:2021yha,Liyanage:2023nds,Giacalone:2023cet,Virta:2024avu,Jaiswal:2025deb}}

We now explain how this scenario also naturally arises from general theoretical considerations~\cite{Eskola:1999fc,Eskola:2001bf,Garcia-Montero:2025hys}. 
We follow the timeline of a nucleus-nucleus collision at ultrarelativistic energies, which consists of several successive stages: 
\begin{itemize}
    \item The collision first produces longitudinally extended tubes of chromoelectric and magnetic fields~\cite{Lappi:2006fp}, analogous to strings~\cite{Artru:1974hr,Andersson:1983ia}. 
    At very early proper time $\tau$, their energy density $\epsilon({\bf x},\tau)$ is independent of $\tau$ and proportional to both thickness functions~\cite{Lappi:2006hq,Nijs:2023yab}:
    \begin{equation}
    \label{step1}
    0^+<\tau<\tau_g: \ \ \epsilon({\bf x},\tau)\propto t_A({\bf x})t_B({\bf x})
    \end{equation}
    \item Fields decay into gluons at a time $\tau_g$. 
    This time is related to the energy density through dimensional analysis, up to logarithmic corrections~\cite{Lappi:2006fp}: $\tau_g\propto\epsilon^{-1/4}\propto (t_At_B)^{-1/4}$. 
    \item $\tau_g$ is typically much smaller than the thermalization time $\tau_h$. 
    In a first approximation, one can neglect interactions, and the longitudinal pressure they generate~\cite{Berges:2020fwq,Jankowski:2020itt}, for $\tau_g<\tau<\tau_h$. 
    During this free-streaming phase~\cite{Blaizot:2019scw}, the energy per unit rapidity is conserved.  
    Since the volume is proportional to $\tau$~\cite{Bjorken:1982qr} (the transverse expansion can be neglected at early times), $\epsilon(\bf x,\tau)\tau$ is constant. 
    Evaluating it at $\tau=\tau_g$, we obtain~\cite{Borghini:2022iym}:
    \begin{equation}
    \label{step2}
    \tau_g<\tau<\tau_h: \ \ \epsilon({\bf x},\tau)\tau\propto (t_A({\bf x})t_B({\bf x}))^{3/4}.  
    \end{equation}
    \item Thermalization occurs at a time which is generically of order $\tau_h\propto T^{-1}$~\cite{Busza:2018rrf}, where  the temperature $T$ is related to $\epsilon(\tau_h)$ by the equation of state. 
    By dimensional analysis, $\epsilon(\tau_h)\propto T^4$ and  $\epsilon(\tau_h)\tau_h\propto T^3$. 
    On the other hand, the entropy density is proportional to $T^3$ so that $s(\tau_h)\tau_h\propto T^2$, and one finally  obtains~\cite{Giacalone:2019ldn}: 
    \begin{align}
    \label{step3}
   s({\bf x},\tau_h)\tau_h &\propto\left(\epsilon({\bf x},\tau_h)\tau_h\right)^{2/3}\nonumber\\
    &\propto (t_A({\bf x})t_B({\bf x}))^{1/2}. 
    \end{align}
    This shows that the entropy density profile at the beginning of the hydrodynamic evolution\footnote{Entropy conservation further implies that $s(\tau)\tau$ remains constant until the transverse expansion sets in~\cite{Ollitrault:2007du}. Therefore, results of the hydrodynamic calculation do not depend strongly on how one chooses $\tau_h$, provided that it is short enough.}  is proportional to $\sqrt{t_At_B}$. 
\end{itemize}
The above modeling is oversimplified, and can be refined at every stage. 
Yet it provides a solid general motivation for choosing $s\propto \sqrt{t_At_B}$. 
We now study the consequences of varying this default initial condition. 

 \section{Simulating the initial state with the generalized \trento{} model} 
 \label{s:trento}

We use the popular \trento{} model of initial conditions~\cite{Moreland:2014oya}, which we briefly describe. 
It starts by sampling the positions of participant nucleons independently  according to  the Monte Carlo Glauber model \cite{Miller:2007ri,Loizides:2014vua}. 
This is typically the most important source of initial-state fluctuations. 
In order to parametrize additional fluctuations, each participant is assigned a random weight.  
The probability of this weight is a gamma distribution with unit mean and variance $1/k$, where the choice of $k$ will be specified below.\footnote{We do not include here fluctuation effects related to the internal structure of the colliding nucleons \cite{Moreland:2018gsh,Kirchner:2025yuo}.} 

The transverse density profile of each participant is modeled  as a two-dimensional Gaussian of width $w_p=0.5$~fm~\cite{Nijs:2022rme,Giacalone:2022hnz}.
The effect of varying the nucleon width is discussed in Appendix~\ref{s:wp}. 
The \trento{}  model then defines the thickness functions of each nucleus $t_A({\bf x})$ and $t_B({\bf x})$ by summing the weighted density profiles of all participants (Appendix~\ref{s:analytic}). 
We parametrize the entropy density as
\begin{equation}
  \label{defnu}
s({\bf x}) \propto  \left(t_A({\bf x})t_B({\bf x})\right)^\nu. 
\end{equation}
The original choice of the \trento{} model is $\nu=0.5$, which is the prescription advocated in Sec.~\ref{s:basics}. 
Here, we generalize it by allowing $\nu$ to vary~\cite{Carzon:2021tif}. 
The motivation is that, in most hydrodynamic simulations, the \trento{} model has been used as an initial condition for the {\it energy\/} density, rather than entropy density~\cite{Bernhard:2019bmu}. 
In order to compensate for this difference, Nijs and van der Schee have generalized the original prescription~\cite{Nijs:2023yab}, and parametrized the energy density as\footnote{More precisely, they define $\epsilon\propto (t_A^p+t_B^p)^{q/p}$, but in practice, $p$ ends up being close to $0$, so that we only study the limit $p\to 0$.}
\begin{equation}
  \label{defq}
\epsilon({\bf x}) \propto  \left(t_A({\bf x})t_B({\bf x})\right)^{q/2}. 
\end{equation}
At high temperature, the equation of state is approximately conformal,  $\epsilon\propto s^{4/3}$, and the correspondence between Eqs.~(\ref{defnu}) and (\ref{defq}) is
\begin{equation}
\label{qvsnu}
q\approx \frac{8}{3}\nu. 
\end{equation}
The default value, corresponding to $\nu=0.5$, is $q=\frac{4}{3}$, but they vary $q$ around this value to study the effect of this parameter in ultracentral collisions~\cite{Nijs:2023bzv}. 
In the same way, we carry out simulations for several values of  $\nu$.

\begin{figure}[t]
   \includegraphics[width=\linewidth]{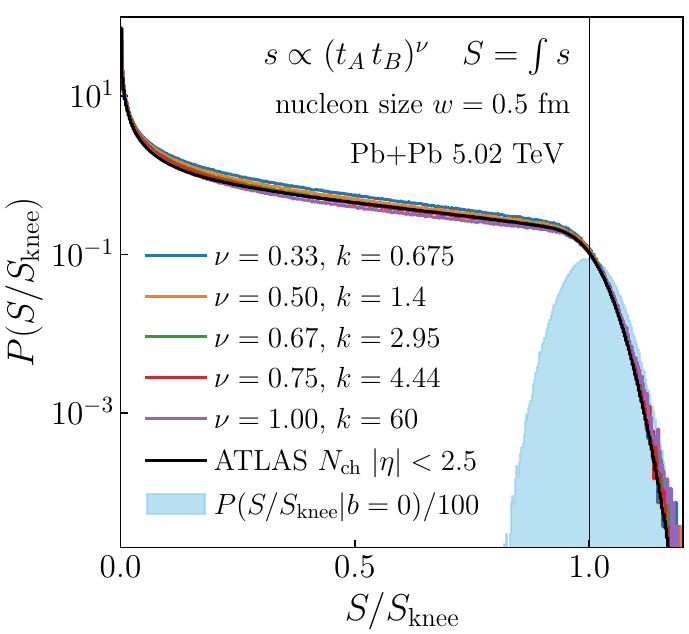}
\caption{\label{fig:histos} 
Histogram of the distribution of the total entropy in a simulation of $10^7$ minimum-bias events for various values of $\nu$, rescaled by its average value at $b=0$, $S_{\rm knee}$. 
For each value of $\nu$, the fluctuation parameter $k$ has a different value determined according to Eq.~(\ref{varS}). 
We also plot the distribution of the charged multiplicity measured by ATLAS~\cite{ATLAS:2024jvf}, rescaled in the same way.
The shaded area displays the distribution of the total entropy for events with $b=0$ for $\nu=0.5$, rescaled by a factor 1/100. It is essentially identical for other values of $\nu$, as a consequence of the constraint (\ref{varS}). 
}
\end{figure}
We now explain how the fluctuation parameter $k$ of the gamma distribution is chosen. 
The idea is that the distribution of the total entropy in minimum-bias events should match the multiplicity distribution measured experimentally, up to a global multiplicity factor. 
This matching is illustrated in Fig.~\ref{fig:histos} for  Pb+Pb collisions at $\sqrt{s_{NN}}=5.02$~TeV. 
It is achieved by first simulating events at $b=0$. 
We define $S_{\rm knee}$ as the mean value of $S$ for these events, and we rescale the entropy by $S_{\rm knee}$. 
The knee of the multiplicity distribution is inferred from data through a  Bayesian reconstruction~\cite{Das:2017ned}. 
We then tune the parameter $k$ of the gamma distribution so as to match the tail of the multiplicity distribution. 
More specifically, we determine $k$ such that the relative variance of the entropy at $b=0$ is identical to that of the multiplicity, which is also inferred from data~\cite{Yousefnia:2021cup,Samanta:2023amp}:
\begin{equation}
\label{varS}
\left(\frac{\sigma_S}{S_{\rm knee}} \right)^2\approx 2\times 10^{-3}, 
\end{equation}
where $\sigma_S$ denotes the standard deviation of $S$ at $b=0$. 
Note that this constraint is not yet enforced in global Bayesian analyses. 
The values of $k$ satisfying Eq.~(\ref{varS}) are displayed in Fig.~\ref{fig:histos}. 
As $\nu$ increases, gamma fluctuations decrease: 
Fluctuations in nucleon position almost suffice for $\nu=1$, while strong gamma fluctuations are needed for $\nu=0.33$ in order to match the tail of the multiplicity distribution. 

\begin{figure*}[t]
   \includegraphics[width=\linewidth]{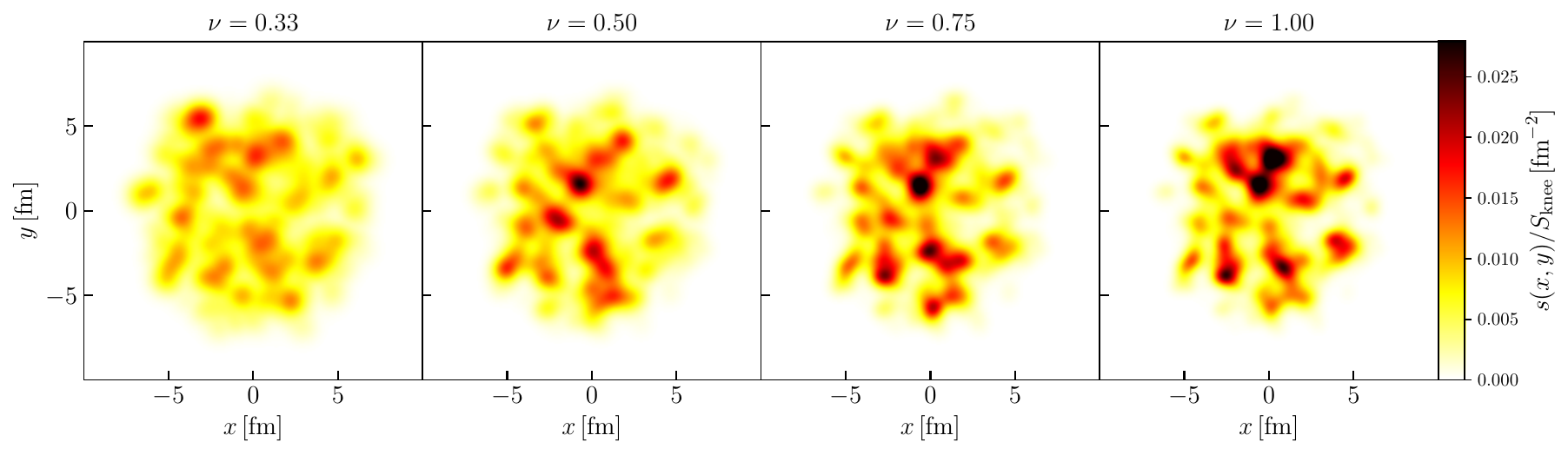}
\caption{\label{fig:profiles} 
Entropy density profiles of collisions with $b=0$, rescaled by a global factor $S_{\rm knee}$ for each $\nu$. 
We vary $\nu$ in Eq.~(\ref{defnu}), keeping the position of nucleons fixed. The gamma fluctuations normalizing each participant nucleon are however sampled independently for each plot (different $k$ parameters).
}
\end{figure*}
Fig.~\ref{fig:profiles} illustrates the variation of the density profile with $\nu$ for a collision at $b=0$, where the nucleon positions are the same for all $\nu$. 
Increasing $\nu$ amplifies density contrasts, as expected from Eq.~(\ref{defnu}). 
Note, however, that changing $\nu$ does not boil down to a nonlinear mapping of the whole profile because of gamma fluctuations, which are sampled independently for each $\nu$. 

\section{Variation of system size with entropy} 
 \label{s:mc}

The main point of this work is to study how the size of the quark-gluon plasma varies with the particle multiplicity or, equivalently, with the total entropy $S$~\cite{Hanus:2019fnc}. 
We define the rms transverse size $R$ of an event from the initial entropy density profile through~\cite{Bozek:2012fw} 
\begin{equation}
\label{defR}
R^2\equiv \frac{1}{S} \int_{\bf x}|{\bf x}|^2 s({\bf x}) -\left|\frac{1}{S} \int_{\bf x}  {\bf x}\, s({\bf x})   \right|^2,
\end{equation}
where the second term in the right-hand side is a recentering correction, which ensures that $R$ is invariant under translations. 

\begin{figure}[t]
   \includegraphics[width=\linewidth]{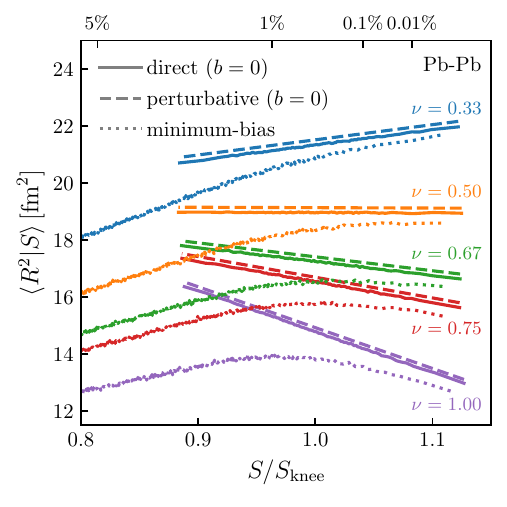}
   \caption{Variation of the average value of $R^2$ (Eq.~(\ref{defR})) with the total entropy  $S$ (Eq.~(\ref{defS})) in Pb+Pb collisions in the \trento{} model, for various values of the exponent $\nu$ in Eq.~(\ref{defnu}). 
   As in Fig.~\ref{fig:histos}, we rescale $S$ by $S_{\rm knee}$.
 Dotted line: minimum-bias events. 
   Solid lines: events with $b=0$. 
Dashed lines: perturbative expression for $b=0$, Eq.~(\ref{R2versusS}). 
}      
        \label{fig:R_vs_S}
\end{figure}
We bin events according to the value of $S$ and evaluate the average value of $R^2$, denoted by $\langle R^2|S\rangle$, in each bin. 
Results are displayed as dotted lines in Fig.~\ref{fig:R_vs_S}, where we only display values of $S/S_{\rm knee}$ above $0.8$, corresponding roughly to the 5\%  most central collisions.  
A first observation is that the size decreases as $\nu$ increases, as could already be guessed from Fig.~\ref{fig:profiles}.  

Let us now look at the variation of $\langle R^2|S\rangle$ with $S$. 
It first increases, reflecting that the impact parameter decreases and the overlap area between the colliding nuclei increases. 
Above the knee, the variation depends on the value of $\nu$: 
The radius keeps increasing for $\nu<0.5$, while it reaches a maximum and then decreases for $\nu>0.5$.  
This is in qualitative agreement with the \textit{Trajectum}   results displayed in Fig.3 (right) of Ref.~\cite{Nijs:2023bzv}, which show that the size keeps increasing for $q=1.05$, corresponding to $\nu\approx 0.39$, and decreases for $q=1.45$, corresponding to $\nu\approx 0.54$. 

In order to understand this behaviour, we carry out a simpler simulation.  
We fix the impact parameter to $b=0$, so that the geometry is frozen, and we simulate $5\times 10^5$ events for each $\nu$. 
The distribution of entropy for these events is a Gaussian centered around the knee~\cite{Das:2017ned}, as illustrated in Fig.~\ref{fig:histos}. 
The size of these central events is displayed as solid lines in Fig.~\ref{fig:R_vs_S}. 
Both figures show that minimum-bias results converge asymptotically to $b=0$ results above the knee. 
Therefore, in order to understand ultracentral collisions, it suffices to understand the case $b=0$.  
Interestingly, for $b=0$, $\langle R^2|S\rangle$ is essentially constant for the default value, $\nu=0.5$. 
This special case is worked out analytically in Appendix~\ref{s:analytic}. 

\section{Relating the variation of the size to microscopic density fluctuations}
\label{s:perturbative}

Focusing on collisions at $b=0$, 
we now relate the variation of the size, $R^2$, with $S$ to the statistical properties of density fluctuations. 

\subsection{Mean density profile}
We first study the mean density profile. 
We denote  by $\kappa_1({\bf x})$ the expectation value of $s({\bf x})$ over all $b=0$ events. 
Its integral is the mean value of $S$ at $b=0$, that is,  $S_{\rm knee}$:
\begin{align}
  \kappa_1({\bf x})&\equiv \langle s({\bf x})\rangle,\nonumber\\
\int_{\bf x}\kappa_1({\bf x}) &=S_{\rm knee}.
  \label{defkappa1}
\end{align}
We choose the origin at the centre of the colliding nuclei. 
Then, $\kappa_1$ depends only on the radial distance $r\equiv |{\bf x}|$ by azimuthal symmetry. 
\begin{figure}[t]
   \includegraphics[width=\linewidth]{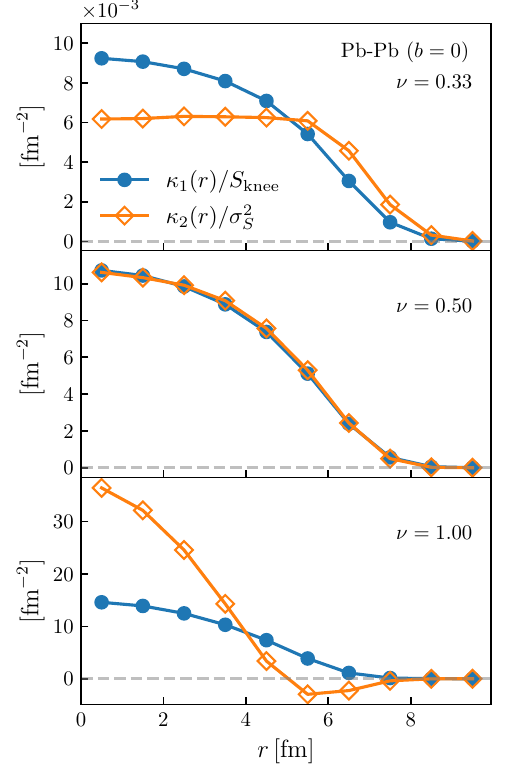}
   \caption{Distribution of the mean entropy density  (circles, Eq.~(\ref{defkappa1})) and of the excess density (squares, Eq.~(\ref{defkappa2})) for three different values of the exponent $\nu$ in Eq.~(\ref{defnu}). 
   Both are normalized in such a way that they integrate to unity over the transverse plane.} 
        \label{fig:kappa}
\end{figure}

Fig.~\ref{fig:kappa} (full symbols) displays $\kappa_1(r)$ for three values of $\nu$, scaled by $S_{\rm knee}$. 
The scaled profile depends weakly on $\nu$. 
The density at the centre $r=0$ mildly increases as a function of $\nu$, implying a sharper average density distribution. 

\begin{figure}[t]
   \includegraphics[width=\linewidth]{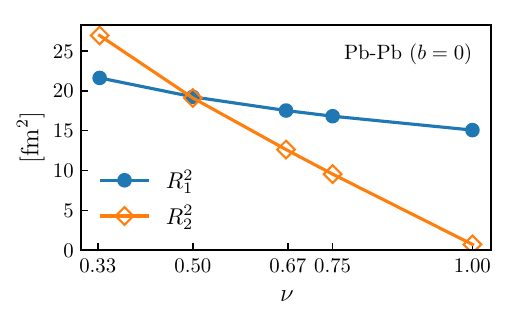}
   \caption{Variation of $R_1^2$ and $R_2^2$, defined by Eqs.~(\ref{defR1}) and (\ref{defR2}), with the exponent $\nu$ in Eq.~(\ref{defnu}).
}      
        \label{fig:r12}
\end{figure}
We denote by $R_1$ the size of the average profile. 
It is obtained by replacing $s({\bf x})$ with $\kappa_1({\bf x})$ in Eq.~(\ref{defR}), where the last  term vanishes by symmetry:
\begin{equation}
\label{defR1}
R_1^2\equiv \frac{1}{S_{\rm knee}} \int_{\bf x} |{\bf x}|^2  \kappa_1({\bf x}).  
\end{equation}
As shown in Fig.~\ref{fig:r12}, $R_1^2$ decreases as $\nu$ increases, in line with the general trend observed in Fig.~\ref{fig:R_vs_S}. 
The decrease of $R_1^2$ with $\nu$ goes along with the increase of the central density $\kappa_1(r=0)$ in Fig.~\ref{fig:kappa}.

\subsection{Fluctuation decomposition}
We now carry out a fluctuation decomposition of the entropy density and of the total entropy: 
\begin{align}
 \label{defdeltas}
s({\bf x})&=  \kappa_1({\bf x})+\delta s({\bf x}),\nonumber \\
S&=S_{\rm knee} +\delta S,
\end{align}
where $\delta s({\bf x})$ is the local fluctuation around the mean, and $\delta S$ its integral over the transverse plane. 
Inserting the decomposition (\ref{defdeltas}) into Eq.~(\ref{defR}) and linearizing in $\delta s({\bf x})$ and $\delta S$, one obtains: 
\begin{equation}
\label{linearize}
R^2= R_1^2+ \frac{1}{S_{\rm knee}}\int_{\bf x}   |{\bf x}|^2 \delta s({\bf x})  -R_1^2\frac{\delta S}{S_{\rm knee}}.
\end{equation}
In order to explain the numerical results in Fig.~\ref{fig:R_vs_S}, we must average this expression over events for a fixed total entropy $S$. 
For this, we need to evaluate the average $\delta s({\bf x})$ at fixed $S$, $\langle \delta s({\bf x})|S \rangle$. 

As we shall see shortly, this quantity is determined by the linear correlation between the entropy density $s({\bf x})$ and the total entropy $S$, which we denote by  $\kappa_2({\bf x})$: 
\begin{equation}
\label{defkappa2}
\kappa_2({\bf x})\equiv \langle \delta s({\bf x})\delta S\rangle=
\int_{{\bf y}}\langle \delta s({\bf x})\delta s({\bf y})\rangle.
\end{equation}
The last equality shows that $\kappa_2({\bf x})$ is the integral of the two point function $\langle \delta s({\bf x})\delta s({\bf y})\rangle$, which is the key quantity that defines event-by-event fluctuations~\cite{Blaizot:2014nia}, over one of the coordinates. 
If one integrates over the remaining coordinate, one obtains the variance of the total entropy: 
\begin{equation}
  \label{sumrule}
\int_{\bf x}\kappa_2({\bf x})= \langle \delta S^2\rangle=\sigma_S^2. 
\end{equation}
In order to evaluate $\langle \delta s({\bf x})|S \rangle$, 
we assume that fluctuations are approximately Gaussian~\cite{Voloshin:2007pc}, which is a good approximation for a collision between large nuclei at fixed impact parameter~\cite{Das:2017ned,Samanta:2023amp}. 
More specifically, we assume that the joint distribution of $\delta s({\bf x})$ and $\delta S$ is a bivariate correlated Gaussian. 
Due to the correlation, the entropy density $s({\bf x})$ is shifted by an amount which is on average proportional to $\delta S$, 
$\langle \delta s({\bf x})|\delta S\rangle=\alpha\delta S$. 
The proportionality constant $\alpha$ is obtained by multiplying with $\delta S$ and integrating over $\delta S$: 
The left-hand side then yields the linear correlation $\langle\delta s({\bf x})\delta S\rangle$. 
One thus obtains: 
\begin{equation}
\label{densityultracentral}
\langle \delta s({\bf x})|S\rangle=
\frac{\langle \delta s({\bf x})\delta S\rangle}{\langle \delta S^2\rangle}\delta S=
\frac{\kappa_2({\bf x})}{\sigma_S^2}\delta S. 
\end{equation}
Using the normalization (\ref{sumrule}), one obtains $\int_{\bf x}\langle \delta s({\bf x})|S\rangle=\delta S$: 
The local fluctuation integrates to $\delta S$, as it should. 

Eq.~(\ref{densityultracentral}) shows that  $\kappa_2({\bf x})$ determines the distribution of the excess density in the transverse plane. 
Its variation is displayed in Fig.~\ref{fig:kappa} for three values of $\nu$. 
For $\nu=0.5$, the distribution of the excess density is exactly the same as that of the mean density, explaining why the radius does not vary (Fig.~\ref{fig:R_vs_S}). 
This result is derived analytically in Appendix~\ref{s:analytic}.
For $\nu<0.5$, the excess density is larger near the edge of the fireball, explaining why the radius increases for ultracentral collisions. 
For $\nu>0.5$, it is the other way around. 
Interestingly, the excess density is even {\it negatively\/} correlated with the total entropy for large $r$. 
The physical mechanism at work in this negative correlation is the fact that the total number of participant cannot exceed the total number of nucleons $2A$.\footnote{This condition is also responsible for the ``binomial suppression" of fluctuations of the net baryon number~\cite{Bzdak:2012ab,STAR:2013gus,Rogly:2018kus,Braun-Munzinger:2023gsd} and of the charged multiplicity~\cite{Roubertie:2025qps}. 
}
Because of this conservation law, more participant nucleons near the edge (hence a larger entropy density) implies fewer in the centre. 
Since most of the entropy is produced in the centre for large $\nu$,  this goes along with a decrease in the total entropy, thus explaining the negative correlation. 

We define a radius $R_2$ associated with the function $\kappa_2({\bf x})$ in the same way as $R_1$  in Eq.~(\ref{defR1}): 
\begin{equation}
\label{defR2}
R_2^2\equiv \frac{1}{\sigma_S^2} \int_{\bf x} |{\bf x}|^2  \kappa_2({\bf x}).  
\end{equation}
We keep the same notation for simplicity, but since $\kappa_2({\bf x})$ can be negative, $R_2^2$ can be negative as well. 
Its variation with $\nu$ is displayed in Fig.~\ref{fig:r12}. 
It is much steeper than that of the average size $R_1^2$. 

We are now in a position to explain the Monte Carlo results of Sec.~\ref{s:mc}. 
We evaluate the mean value of Eq.~(\ref{linearize}) at fixed $S$ using Eq.~(\ref{densityultracentral}). 
We express the result in terms of $R_2$ using Eq.~(\ref{defR2}), and we obtain: 
\begin{equation}
\label{R2versusS}
\langle R^2|S\rangle=R_1^2+\left(R_2^2-R_1^2  \right)\frac{\delta S}{S_{\rm knee}}.
\end{equation}
The average value of $R^2$ for $S=S_{\rm knee}$ is $R_1^2$, as one can check by comparing Fig.~\ref{fig:R_vs_S} with Fig.~\ref{fig:r12}. 
The radius increases or decreases with $S$ depending on whether $R_2$ is larger or smaller than $R_1$. 
The perturbative result (\ref{R2versusS}) is in excellent agreement with the Monte Carlo calculation, as illustrated in Fig.~\ref{fig:R_vs_S}, which is essentially our main result.
For $\nu=0.5$, $R_1=R_2$ and $\langle R^2|S\rangle$ is independent of $S$.

\section{Increase of $\langle p_T\rangle$ in ultracentral collisions}
\label{s:trajectum}

We now discuss potential experimental implications of our finding. 
We first describe how the mean $p_T$ of outgoing particles, $\langle p_T\rangle$, depends on the initial density profile in hydrodynamic simulations at $b=0$.

It has been observed that in hydrodynamics, $\langle p_T\rangle$ is proportional to an effective temperature $T_{\rm eff}$ which is defined from the energy $E$ and entropy $S$ of the fluid per unit space-time rapidity at freeze-out through the equations~\cite{Gardim:2019xjs}: 
\begin{align}
\label{effectivehydro}
E&=\epsilon(T_{\rm eff}) V_{\rm eff},\nonumber\\
S&=s(T_{\rm eff}) V_{\rm eff},
\end{align}
where $\epsilon$ and $s$ denote the energy density and entropy density, which are related to the temperature through the equation of state used in the hydrodynamic simulation. 
The proportionality between $\langle p_T\rangle$ and $T_{\rm eff}$ is approximate and cannot be derived rigorously~\cite{Gavassino:2025bts}, so that it eventually boils down to a numerical observation. 
One of its remarkable features is that it is extremely robust with respect to variations of the transport coefficients, which are the main sources of uncertainties in hydrodynamic simulations~\cite{JETSCAPE:2020mzn}. 

The relevant characteristic length for the hydrodynamic expansion is the initial transverse size $R$. 
For dimensional reasons, $V_{\rm eff}$ is proportional to $R^3$: 
The volume in a rapidity slice $dy$ is the product of a transverse area, proportional to $R^2$, by a longitudinal size $\tau dy$, where $\tau$ is the time at which longitudinal cooling stops (after which that both energy and entropy are approximately conserved), which is also proportional to $R$ for dimensional reasons. 
Hydrodynamic simulations at $b=0$ with fluctuating initial conditions confirm a strong linear correlation between $V_{\rm eff}$  and  $R^3$, with~\cite{Gardim:2020sma}
\begin{equation}
\label{veff}
V_{\rm eff}\approx 0.9\pi (R\sqrt 2)^3.
\end{equation}
We now explain how the speed of sound $c_s$ can be related to observables~\cite{Gardim:2019brr,CMS:2024sgx}. 
By definition of $c_s$,
\begin{equation}
\label{defcs}
d\ln T_{\rm eff}=c_s^2(T_{\rm eff})\,  d\ln s(T_{\rm eff}).
\end{equation}
We can generally decompose the variations on both sides as:
\begin{align}
\label{decomposition}
d\ln T_{\rm eff}&=d\ln\langle p_T\rangle-d \ln\left(\frac{\langle p_T\rangle}{T_{\rm eff}}\right)\nonumber\\
d\ln  s(T_{\rm eff})&= d\ln S-d\ln\left(\frac{V_{\rm eff}}{R^3}\right)-d\ln R^3. 
\end{align}
Assuming that $\langle p_T\rangle/T_{\rm eff}$ and $V_{\rm eff}/R^3$ are constant, and  $S\propto N_{ch}$, Eqs.~(\ref{defcs}) and (\ref{decomposition}) give: 
\begin{equation}
\label{nonconstantR}
d\ln\langle p_T\rangle=c_s^2(T_{\rm eff})\, (d\ln N_{ch}-d\ln R^3).
\end{equation}
For constant $R$, this equation reduces to 
\begin{equation}
\label{constantR}
d\ln\langle p_T\rangle=c_s^2(T_{\rm eff})\, d\ln N_{ch}. 
\end{equation}
An interesting recent development is that this equation holds very precisely in hydrodynamics, even though $\langle p_T\rangle/T_{\rm eff}$ and $V_{\rm eff}/R^3$ are not strictly constant~\cite{Gavassino:2025bts}. 
It turns out that both ratios increase mildly with $N_{ch}$~\cite{Gardim:2024zvi}, but these increases satisfy 
\begin{equation}
\label{coincidence}
d\ln\left(\frac{\langle p_T\rangle}{T_{\rm eff}}\right)\approx c_s^2(T_{\rm eff})\, d\ln \left(\frac{V_{\rm eff}}{R^3}\right),
\end{equation}
so that their contributions mutually cancel, and Eq.~(\ref{constantR}) still holds. 
While there is no rigorous mathematical derivation of  Eq.~(\ref{constantR}), the difference between the two sides is smaller than $0.01$ for a broad range of colliding energies, freeze-out temperatures~\cite{Gavassino:2025bts}, and equations of state~\cite{Gardim:2024zvi}, so that it is unlikely to be a mere coincidence. 
In any case, it is a useful numerical observation which is firmly  established for smooth initial conditions and equations of state resembling that of QCD.\footnote{Large violations occur if the equation of state has abrupt variations around $T_{\rm eff}$~\cite{Gavassino:2025bts}, which is not expected for QCD.} 
We will assume that it remains valid in the presence of initial-state fluctuations \cite{Gardim:2020sma,Sun:2024zsy,Mu:2025gtr}, which should eventually be confirmed by a dedicated study. 

We now derive the modifications which apply if $R$ varies with the initial entropy $S$, as studied in this paper. 
If Eq.~(\ref{constantR}) is valid for constant $R$, then Eq.~(\ref{nonconstantR}) holds more generally if $R$ varies, as the dependence on $R$ merely follows from dimensional analysis. 
We write $\delta\ln R^3=(3/2)\delta R^2/R_1^2$, 
where $R_1^2$ is the mean value of $R^2$ as shown in Sec.~\ref{s:perturbative}, and 
$\delta R^2=R^2-R_1^2$ is the variation around the mean. 
Using Eq.~(\ref{R2versusS}), and assuming that the entropy is proportional to $N_{ch}$, Eq.~(\ref{nonconstantR}) gives: 
\begin{equation}
\label{eq3pt}
d\ln\langle p_T\rangle=
c_s^2(T_{\rm eff})  
\left[1-\frac{3}{2}\left(\frac{R_2^2}{R_1^2}-1\right)\right] d\ln N_{ch}. 
\end{equation}
We define the slope parameter as 
\begin{align}
{\rm slope}\equiv
\frac{\delta\ln\langle p_T\rangle}{\delta\ln N_{ch}}
=c_s^2(T_{\rm eff})  
\left(1-\frac{3}{2}\left(\frac{R_2^2}{R_1^2}-1\right)\right). 
\label{slope}
\end{align}
The assumption of constant size~\cite{Gardim:2024zvi,Gavassino:2025bts} corresponds to $R_2=R_1$, and one recovers the result that the slope is $c_s^2(T_{\rm eff})$. 
Our formula thus generalizes this case by allowing for a volume variation driven by local density fluctuations and the relative difference between $R_2^2$ and $R_1^2$. 

\begin{figure}[t]
   \includegraphics[width=\linewidth]{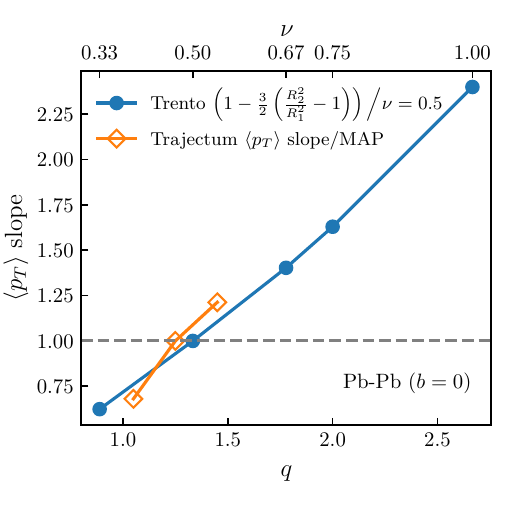}
   \caption{
   Slope defined by Eq.~(\ref{slope}), scaled by the default value $c_s^2$. 
   Orange squares: Numerical results using the \textit{Trajectum} framework~\cite{Nijs:2023bzv} as a function of the exponent $q$ in Eq.~(\ref{defq}).
   Blue circles: Our perturbative result (\ref{eq3pt}) as a function of the exponent $\nu$ in Eq.~(\ref{defnu}). 
   We assume for simplicity that $q$ and $\nu$ are related through Eq.~(\ref{qvsnu}). 
}      
        \label{fig:slopes}
\end{figure}

In Fig.~\ref{fig:slopes}, we compare our results with \textit{Trajectum} results, obtained from full hydrodynamic simulations~\cite{Nijs:2023bzv}. 
We assume that the initial energy density profile of \textit{Trajectum} is related to entropy density profile in our calculation through Eq.~(\ref{qvsnu}). Note that this correspondence is approximate. 
We evaluate the slope in \textit{Trajectum} using results displayed in Fig.3 (left) of Ref.~\cite{Nijs:2023bzv}, where we extract the slope of the $\langle p_T \rangle$ curve from the largest values of $N_{ch}$, corresponding to ultracentral collisions. 

The slope of the hydrodynamic calculation deviates significantly from    Eq.~(\ref{slope}) due to a number of effects, which are listed in Ref.~\cite{Nijs:2023bzv}.  
Some of these effects have been quantitatively evaluated, in particular that of transverse momentum cuts in the detector acceptance~\cite{Parida:2024ckk}
and that of statistical (Poisson) fluctuations of $N_{ch}$~\cite{Gardim:2024zvi}.\footnote{
Cuts in pseudorapidity also have a non-trivial effect~\cite{SoaresRocha:2024drz,Gardim:2024zvi,Gavassino:2025bts}, but a smaller one.}
Both amount to a global multiplicative factor of the slope, so that the relative modification due to the volume variation is the same.

Therefore, in order to compare our results with \textit{Trajectum}, we take ratios of slopes, with respect to a default scenario where the size is essentially constant. 
The increase of the slope as a function of $q$ in \textit{Trajectum} calculations is similar to the increase as a function of $\nu$ in our initial-state simulations, which shows that our simple modeling captures the origin of the effect.  
More quantitative comparisons with a precise matching between initial conditions and final observables are left for future work.

Equation~(\ref{slope}) also rules the increase of $\langle p_T\rangle$ as a function of collision energy at fixed centrality. 
By comparing Pb+Pb results at $\sqrt{s_{NN}}=2.76$~TeV and $5.02$~TeV, it has been shown that the value $c_s^2(T_{\rm eff})$ inferred from this relation is compatible with that calculated in lattice QCD~\cite{Gardim:2019xjs}, assuming that the volume does not vary with energy, i.e., setting $R_2=R_1$ in Eq.~(\ref{slope}). 

With all these considerations in mind, we would like to emphasize that that the main message of this work holds irrespective of our ability of relating $c_s^2$ and $\langle p_T\rangle$ through simple equations. 
After all, multiple probes and temperature determinations show that the equation of state relevant for heavy-ion collisions is that of hot QCD. 
Similarly, the statement that the ultracentral $\langle p_T\rangle$ slope should correlate with the multiplicity dependence of the initial QGP volume should naturally be true in general. 
Consequently, we expect that it will eventually be possible to use precision measurements of the increase of $\langle p_T\rangle$ in ultracentral collisions to place quantitative bounds on the relative difference between $R_2^2$ and $R_1^2$. As we have demonstrated, this will offer new, non-trivial information on the distribution of entropy density fluctuations across the transverse plane.

\section{Conclusions}

We have shown that the variation of the size of the quark-gluon plasma in ultracentral collisions depends on how the excess density is distributed in the transverse plane. 
The statistics of energy fluctuations at the time when the two nuclei interact is modified through a nonlinear evolution by the pre-equilibrium dynamics, and leads to a scaling of the initial entropy density of the type $\sqrt{t_At_B}$, making the total entropy proportional to the mass number of the colliding ions in the ultra-central limit ($t_A=t_B$), which is well motivated by experimental results. Our main finding is that, if the multiplicity is proportional to the number of nucleons, then the excess entropy density is distributed in the same way as the average density in the transverse plane. 

This leaves little room for a variation of the volume with the multiplicity. However, as measurements of $\langle p_T \rangle$ in ultra-central collisions are extremely precise, it should be feasible to make precise statements in this regard based on existing data. It would be important, in particular, to repeat this analysis in ultra-central light-ion collisions \cite{ATLAS:2025nnt,ALICE:2025luc,CMS:2025tga}. Although highly consistent with hydrodynamic expectations, in such systems pre-equilibrium corrections are expected to play a more important role (especially at RHIC energies \cite{STAR:2025ivi}), potentially modifying the statistics of fluctuations of the entropy density field.

We emphasize that verifying the scaling of the entropy density with the nuclear mass numbers through such measurements can have important consequences for the study of nuclear structure at colliders \cite{Jia:2022ozr,STAR:2024wgy,Giacalone:2025vxa}. If the entropy density is such that $s({\bf x})\propto t({\bf x})$ in the ultracentral limit, this implies that the collision process itself does not generate any new spatial correlations in the transverse plane \cite{Giacalone:2023hwk}. Therefore, if the increase of $\langle p_T\rangle$ in ultracentral collisions confirms that the volume of the quark-gluon plasma does not increase or decrease as a function of the multiplicity, it will strengthen the argument that fluctuations in ultracentral collisions are essentially those coming from one nucleus alone. This in turn entails a straightforward relation between multi-particle correlation observables in symmetric heavy-ion collisions and many-body correlations in the ground state of the nucleus~\cite{Duguet:2025hwi}, as illustrated in Appendix~\ref{s:analytic} for two-body correlations. This will allow high-energy colliders to probe many-body properties of nuclear ground states in unprecedented detail.

\begin{acknowledgments}
We thank Jean-Paul Blaizot and Aleksas Mazeliauskas for numerous discussions, and Govert Nijs and Wilke van der Schee for sharing Trajectum results. 
 F.Z. acknowledges support by the DFG through Emmy Noether Programme (project number 496831614) and CRC 1225 ISOQUANT (project number 27381115) as well as the state of Baden-Württemberg through bwHPC.
\end{acknowledgments}

\appendix

\section{Why $s\propto\sqrt{t_At_B}$ is special}
\label{s:analytic}

In this Appendix, we derive approximate analytic expressions of density correlations in the \trento{} model for Pb+Pb collisions at  $b=0$. 
This specific case is simpler  for two reasons: 
\begin{itemize}
\item Target and projectile play symmetric roles at each point ${\bf x}$. 
\item Almost all the nucleons (98\% on average in Monte Carlo Glauber calculations) participate in the collision. We make the approximation that {\it all\/} nucleons participate, which is the case at asymptotically high energies.  
\end{itemize}
 We start from the expression of the entropy density, Eq.~(\ref{defnu}), in which we carry out a standard  fluctuation decomposition: 
\begin{equation}
\label{deltata}
t_{A,B}({\bf x})= \langle t_A({\bf x})\rangle + \delta t_{A,B}({\bf x}),
\end{equation}
where local symmetry between target and projectile implies $\langle t_A({\bf x})\rangle= \langle t_B({\bf x})\rangle$.
Inserting in Eq.~(\ref{defnu}), where we omit the global multiplicative constant, and linearizing in the fluctuations, one obtains 
\begin{align}
\kappa_1({\bf x})&= \langle t_A({\bf x})\rangle^{2\nu}\nonumber\\
\delta s({\bf x})&=\nu  \langle t_A({\bf x})\rangle^{2\nu-1}\left(\delta t_A({\bf x})+\delta t_B({\bf x})\right). 
\end{align}
One immediately sees why $\nu=\frac{1}{2}$ is special, as these expressions simplify to: 
\begin{align}
\kappa_1({\bf x})&= \langle t_A({\bf x})\rangle\nonumber\\
\delta s({\bf x})&=\frac{1}{2}\left(\delta t_A({\bf x})+\delta t_B({\bf x})\right). 
\end{align}
These equations show that both the mean value and the fluctuation are linear in the thickness function for this specific choice of $\nu$. 
We assume $\nu=\frac{1}{2}$ from now on. 

The two point-function $\langle \delta s({\bf x})\delta s({\bf y})\rangle$, which will be needed in order to evaluate $\kappa_2({\bf x})$, contains four terms. 
Since the two nuclei are independent, the cross correlation vanishes:
$\langle\delta t_A({\bf x})\delta t_B({\bf y})\rangle=0$. 
This simplification arises because we have assumed that all the nucleons are participants. 
Otherwise, the condition that a nucleon participates induces correlations between the two colliding nuclei, which are dubbed ``twin correlations''~\cite{Blaizot:2014wba}. 
The only nonvanishing terms are $\langle \delta t_A({\bf x})\delta t_A({\bf y})\rangle$ and  $\langle \delta t_B({\bf x})\delta t_B({\bf y)}\rangle$ which are equal by symmetry. 
We thus obtain: 
\begin{equation}
  \label{2points}
 \langle \delta s({\bf x})\delta s({\bf y})\rangle=\frac{1}{2} \langle \delta t_A({\bf x})\delta t_A({\bf y})\rangle. 
\end{equation}
This equation shows that the two-point correlation of the entropy density profile is simply related to that of a single nucleus~\cite{Duguet:2025hwi}. 

We have related the statistical properties of the entropy density to those of the thickness function $t_A({\bf x})$. 
In order to evaluate them, we go back to the definition~\cite{Moreland:2014oya}: 
\begin{equation}
\label{trentota}
  t_A({\bf x})=\sum_{i=1}^{A}w_i\rho_{\rm p}({\bf x}-{\bf x}_i),
\end{equation}
where the sum runs over all nucleons, $A=208$ for Pb, and
\begin{itemize}
\item
$w_i$ are independent weights sampled according to a gamma distribution, which satisfies
\begin{align}
\langle w\rangle &= 1\nonumber\\
\langle w^2\rangle &= 1+\frac{1}{k}. 
\label{gamma}
\end{align}
\item 
${\bf x}_i$ is the transverse position of nucleon $i$,  
\item
$\rho_{\rm p}({\bf x})$ is the normalized ``nucleon profile''. 
\end{itemize}
We make the approximation that the nucleon size is much smaller than the nuclear radius, and treat the  profile as a Dirac peak: $\rho_{\rm p}({\bf x})\approx \delta({\bf x})$. 
The effect of the nucleon size is discussed in Appendix~\ref{s:wp}. 

We now evaluate the one- and two-point functions of $t_A({\bf x})$. 
The probability distribution of ${\bf x}_i$ is $T_A({\bf x}_i)/A$, where $T_A({\bf x})$ denotes the usual thickness function of the nucleus in an optical Glauber calculation~\cite{Miller:2007ri}, that is, the integral of the nucleon density over the longitudinal coordinate. 
Since there are $A$ identical terms, the mean value of $t_A({\bf x})$ is:
\begin{equation}
  \label{meantrento2}
  \kappa_1({\bf x})=\langle t_A({\bf x})\rangle=\int_{\bf y} T_A({\bf y}) \delta({\bf x}-{\bf y})=T_A({\bf x}),
\end{equation}
where we have used $\langle w_i\rangle=1$. 

We now evaluate the two-point function.
Eq.~(\ref{trentota}) gives
\begin{equation}
  \label{tata}
  t_A({\bf x})t_A({\bf y})=\sum_{i=1}^{A}\sum_{j=1}^{A} w_iw_j
 \delta({\bf x}-{\bf x}_i)\delta({\bf y}-{\bf x}_j).
\end{equation}
There are $A^2$ terms, which consist of $A(A-1)$ non-diagonal terms with $i\not= j$ and $A$ diagonal terms.
Since we consider spherical $^{208}$Pb nuclei, the nucleons are sampled independently, such that  ${\bf x}_i$ and ${\bf x}_j$ are independent variables for $i\not= j$ (unless an excluded volume is implemented in the Glauber calculation~\cite{Luzum:2023gwy}). 
We average over events, separating the diagonal and non-diagonal terms, and using Eqs.~(\ref{gamma}): 
\begin{align}
  \label{avtata}
 \langle  t_A({\bf x})t_A({\bf y})\rangle &=\left(1+\frac{1}{k}\right)T_A({\bf x}) \delta({\bf x}-{\bf y})\nonumber\\
&+ \left(1-\frac{1}{A}\right) T_A({\bf x})T_A({\bf y}).
\end{align}
Using Eqs.~(\ref{deltata}), (\ref{meantrento2}) and (\ref{avtata}), we obtain the following expression of the density-density correlation (\ref{2points}): 
\begin{align}
  \label{2point}
  \langle \delta s({\bf x})\delta s({\bf y})\rangle&=
\frac{1}{2}\left( \langle  t_A({\bf x})t_A({\bf y})\rangle - \langle  t_A({\bf x})\rangle \langle t_A({\bf y})\rangle\right)\nonumber\\
  &= \frac{1}{2}\left(1+\frac{1}{k}\right)T_A({\bf x}) \delta({\bf x}-{\bf y})\nonumber\\
  &-\frac{1}{2A}T_A({\bf x}) T_A({\bf y}).
\end{align}
It consists of a positive short-range part, $\delta({\bf x}-{\bf y})$, and a negative long-range part induced by the condition that the total number of nucleons is fixed. 
In the case $\nu=1$, a similar negative long-range correlation is present, whose analytic expression could easily be obtained in the same way. 
It explains the negative values of $\kappa_2(r)$ observed in the bottom panel of Fig.~\ref{fig:kappa} and discussed in Sec.~\ref{s:perturbative}.  

We finally integrate Eq.~(\ref{2point})  over ${\bf y}$ to obtain  $\kappa_2({\bf x})$, defined by Eq.~(\ref{defkappa2}): 
\begin{equation}
\label{kappa2trento}
\kappa_2({\bf x})=\int_{\bf y} \langle \delta s({\bf x})\delta s({\bf y})\rangle=
\frac{1}{2k}T_A({\bf x}), 
\end{equation}
where we have used $\int_{\bf y} T_A({\bf y})=A$. 
Comparing with Eq.~(\ref{meantrento2}), one sees that 
$\kappa_2({\bf x})$ and $\kappa_1({\bf x})$ are proportional to one another, as observed in Fig.~\ref{fig:kappa}. 
According to Eq.~(\ref{densityultracentral}), this means that excess density in ultracentral collision is distributed in the same way as the mean density. 
This explains why the radius is independent of the size, as observed in Fig.~\ref{fig:R_vs_S} for $\nu=0.5$.

As a byproduct of this calculation, we finally  derive an analytic expression of the variance of the total entropy. 
Using Eqs.~(\ref{defkappa1}), (\ref{sumrule}), (\ref{meantrento2}) and (\ref{kappa2trento}), one obtains
\begin{equation}
\label{sigmaS}
\frac{\sigma_S^2}{S_{\rm knee}^2}=\frac{\int_{\bf x} \kappa_2({\bf x})}{\left(\int_{\bf x} \kappa_1({\bf x})\right)^2}=\frac{1}{2kA}.
\end{equation}
The value of $k$ that satisfies the constraint from ATLAS data, Eq.~(\ref{varS}), is $k\approx 1.2$, in reasonable agreement with the value $k=1.4$ reported in Fig.~\ref{fig:histos}. 
This equation also shows that the gamma fluctuations implemented in the \trento{} model via the parameter $k$ are solely responsible for the fluctuations of the total entropy for $\nu=0.5$. 
In the limit $k\to\infty$, the weights $w_j$ in Eq.~(\ref{trentota}) are all equal to unity, and the only remaining fluctuations are those of the positions of nucleons. 
In this limit, entropy fluctuations vanish, which means that fluctuations in nucleon positions do not contribute to entropy fluctuations to leading order. 
It is quite a remarkable result, as these fluctuations produce sizable fluctuations in other observables, such as initial anisotropies~\cite{PHOBOS:2006dbo,Alver:2010gr}. 

\section{Influence of the nucleon width}
\label{s:wp}

In the \trento{} model, the nucleon width $w_p$ enters the profile function associated with each participant in  (\ref{trentota}), which is modeled as a Gaussian~\cite{Moreland:2014oya}: 
\begin{equation}
  \label{gaussianprofile}
\rho_p({\bf x})=\frac{1}{2\pi w_p^2}\exp\left(-\frac{{\bf x}^2}{2w_p^2}\right), 
\end{equation}

Within the approximations made in Appendix~\ref{s:analytic}, $w_p$ does not enter any of our results. 
In this Appendix, we illustrate the effect of the nucleon width by evaluating more precisely the average density profile $\kappa_1({\bf x})$, taking into account terms of order 2 in the fluctuations, which have been neglected so far.

Using Eq.~(\ref{deltata}), we obtain
\begin{align}
\label{delta2}
t_A({\bf x})^{1/2}&=\left(T_A({\bf x})+\delta t_A({\bf x})\right)^{1/2} \nonumber\\
&=T_A({\bf x})^{1/2}\left(1+\frac{1}{2}\frac{\delta t_A({\bf x})}{T_A({\bf x})}-\frac{1}{8}\frac{\delta t_A({\bf x})^2}{T_A({\bf x})^2}       \right).
\end{align}
Carrying out a similar decomposition of $t_B({\bf x})$, inserting into Eq.~(\ref{defnu}) with $\nu=\frac{1}{2}$, and averaging over events, we obtain the following expression of the mean entropy density: 
\begin{equation}
\label{delta2av}
\kappa_1({\bf x})=
\langle t_A({\bf x})\rangle-\frac{1}{4} 
\frac{\langle\delta t_A({\bf x})^2\rangle}{T_A(x)}, 
\end{equation}
where we have used the independence between the fluctuations in the two nuclei $\langle \delta t_A({\bf x})\delta t_B({\bf x})\rangle=0$. 
The correction induced by fluctuations is proportional to the  variance of the local density, $\langle\delta t_A({\bf x})^2\rangle$. 

\begin{figure}[t]
   \includegraphics[width=\linewidth]{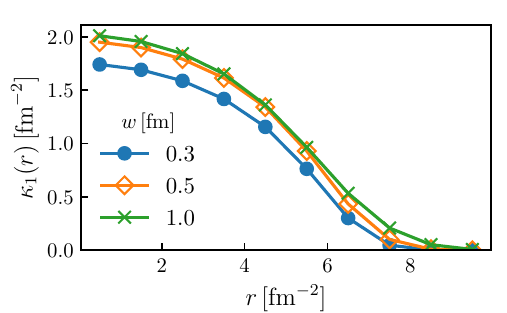}
   \caption{Dependence of the average entropy density $\kappa_1(r)$ on the nucleon radius in the \trento{} model. 
   The entropy density is defined by Eq.~(\ref{defnu}), where $\nu=\frac{1}{2}$ and the proportionality factor has been set to unity. The thickness functions $t_{A,B}({\bf x})$ are defined by Eq.~(\ref{trentota}). Results for Pb+Pb collisions at $b=0$ are displayed for three different values of the nucleon width $w_p$. 
}      
        \label{fig:wp}
\end{figure}

This variance is evaluated using Eq.~(\ref{tata}), in which we set ${\bf y}={\bf x}$, and where we replace the Dirac peaks by the actual density profile $\rho_p({\bf x})$ of the nucleon. 
The dominant contribution comes from the diagonal terms with $i=j$ and one obtains, after averaging over events: 
\begin{align}
\langle\delta t_A({\bf x})^2\rangle&=\langle w_i^2\rangle T_A({\bf x}) \int_{\bf z} \rho_p({\bf z})^2 \nonumber\\
&=\left(1+\frac{1}{k}\right)T_A({\bf x})\frac{1}{4\pi w_p^2},
\end{align}
where, in the last equality,  we have used Eqs.~(\ref{gamma}) and (\ref{gaussianprofile}). 
Inserting into Eq.~(\ref{delta2av}), and using (\ref{meantrento2}), we finally obtain: 
\begin{equation}
\label{meansbetter}
\kappa_1(x)=T_A(x)-\frac{1+\frac{1}{k}}{16\pi w_p^2}. 
\end{equation}
The net effect of fluctuations is that they shift the mean entropy density by a negative additive constant. [Note that the mean entropy density must be positive, so that the validity of Eq.~(\ref{meansbetter}) breaks down for large $r$.]
This shift is inversely proportional to $w_p^2$, which can be understood as follows: 
A smaller nucleon has a smaller probability of colliding with other nucleons, leading to a decrease in the density.  
Fig.~\ref{fig:wp} displays numerical results for three different values of $w_p$. 
Eq.~(\ref{meansbetter}) predicts that $\kappa_1(r)$ is smaller by $0.1$~fm$^{-2}$ for $w_p=0.5$~fm than for $w_p=1$~fm, and by $0.24$~fm$^{-2}$ for $w_p=0.3$~fm than for $w_p=0.5$~fm, in fair agreement with the numerical results.


\begin{thebibliography}{99}
\bibitem{CMS:2013bza}
S.~Chatrchyan \textit{et al.} [CMS],
JHEP \textbf{02} (2014), 088
[arXiv:1312.1845 [nucl-ex]].

\bibitem{Luzum:2012wu}
M.~Luzum and J.~Y.~Ollitrault,
Nucl. Phys. A \textbf{904-905} (2013), 377c-380c
[arXiv:1210.6010 [nucl-th]].

\bibitem{Shen:2015qta}
C.~Shen, Z.~Qiu and U.~Heinz,
Phys. Rev. C \textbf{92} (2015) no.1, 014901
[arXiv:1502.04636 [nucl-th]].

\bibitem{Das:2017ned}
S.~J.~Das, G.~Giacalone, P.~A.~Monard and J.~Y.~Ollitrault,
Phys. Rev. C \textbf{97} (2018) no.1, 014905
[arXiv:1708.00081 [nucl-th]].

\bibitem{Samanta:2023amp}
R.~Samanta, S.~Bhatta, J.~Jia, M.~Luzum and J.~Y.~Ollitrault,
Phys. Rev. C \textbf{109} (2024) no.5, L051902
[arXiv:2303.15323 [nucl-th]].

\bibitem{Liu:2022kvz}
L.~M.~Liu, C.~J.~Zhang, J.~Zhou, J.~Xu, J.~Jia and G.~X.~Peng,
Phys. Lett. B \textbf{834} (2022), 137441
[arXiv:2203.09924 [nucl-th]].

\bibitem{Zhang:2025voj}
H.~Zhang, A.~Akridge, C.~J.~Horowitz, J.~Liao and H.~Xing,
[arXiv:2510.07816 [nucl-th]].

\bibitem{Blaizot:2014nia}
J.~P.~Blaizot, W.~Broniowski and J.~Y.~Ollitrault,
Phys. Lett. B \textbf{738} (2014), 166-171
[arXiv:1405.3572 [nucl-th]].

\bibitem{Giacalone:2023hwk}
G.~Giacalone,
Eur. Phys. J. A \textbf{59} (2023) no.12, 297
[arXiv:2305.19843 [nucl-th]].

\bibitem{Mehrabpour:2025ogw}
H.~Mehrabpour,
[arXiv:2506.12673 [nucl-th]].

\bibitem{Duguet:2025hwi}
T.~Duguet, G.~Giacalone, S.~Jeon and A.~Tichai,
Phys. Rev. Lett. \textbf{135} (2025) no.18, 182301
[arXiv:2504.02481 [nucl-th]].

\bibitem{Liu:2025uks}
Q.~Liu, H.~Mehrabpour and B.~N.~Lu,
[arXiv:2509.00315 [nucl-th]].

\bibitem{Yousefnia:2021cup}
K.~V.~Yousefnia, A.~Kotibhaskar, R.~Bhalerao and J.~Y.~Ollitrault,
Phys. Rev. C \textbf{105} (2022) no.1, 014907
[arXiv:2108.03471 [nucl-th]].

\bibitem{Pepin:2022jsd}
M.~Pepin, P.~Christiansen, S.~Munier and J.~Y.~Ollitrault,
Phys. Rev. C \textbf{107} (2023) no.2, 024902
[arXiv:2208.12175 [nucl-th]].

\bibitem{CMS:2024sgx}
A.~Hayrapetyan \textit{et al.} [CMS],
Rept. Prog. Phys. \textbf{87} (2024) no.7, 077801
[arXiv:2401.06896 [nucl-ex]].

\bibitem{ATLAS:2024jvf}
G.~Aad \textit{et al.} [ATLAS],
Phys. Rev. Lett. \textbf{133} (2024) no.25, 252301
[arXiv:2407.06413 [nucl-ex]].

\bibitem{ALICE:2025rtg}
I.~J.~Abualrob \textit{et al.} [ALICE],
JHEP \textbf{11} (2025), 076
[arXiv:2506.10394 [nucl-ex]].

\bibitem{Gavassino:2025bts}
L.~Gavassino, H.~Hirvonen, J.~F.~Paquet, M.~Singh and G.~Soares Rocha,
Phys. Rev. C \textbf{112} (2025) no.5, 054903
[arXiv:2503.20765 [hep-ph]].

\bibitem{Gardim:2024zvi}
F.~G.~Gardim, A.~V.~Giannini and J.~Y.~Ollitrault,
Phys. Lett. B \textbf{856} (2024), 138937
[arXiv:2403.06052 [nucl-th]].

\bibitem{Gardim:2019brr}
F.~G.~Gardim, G.~Giacalone and J.~Y.~Ollitrault,
Phys. Lett. B \textbf{809} (2020), 135749
[arXiv:1909.11609 [nucl-th]].

\bibitem{Ollitrault:2007du}
J.~Y.~Ollitrault,
Eur. J. Phys. \textbf{29} (2008), 275-302
[arXiv:0708.2433 [nucl-th]].

\bibitem{Nijs:2023bzv}
G.~Nijs and W.~van der Schee,
Phys. Lett. B \textbf{853} (2024), 138636
[arXiv:2312.04623 [nucl-th]].

\bibitem{Sun:2024zsy}
J.~A.~Sun and L.~Yan,
Phys. Lett. B \textbf{866} (2025), 139507
[arXiv:2407.05570 [nucl-th]].

\bibitem{Moreland:2014oya}
J.~S.~Moreland, J.~E.~Bernhard and S.~A.~Bass,
Phys. Rev. C \textbf{92} (2015) no.1, 011901
[arXiv:1412.4708 [nucl-th]].

\bibitem{Nijs:2023yab}
G.~Nijs and W.~van der Schee,
[arXiv:2304.06191 [nucl-th]].

\bibitem{Bjorken:1982qr}
J.~D.~Bjorken,
Phys. Rev. D \textbf{27} (1983), 140-151

\bibitem{Eremin:2003qn}
S.~Eremin and S.~Voloshin,
Phys. Rev. C \textbf{67} (2003), 064905
[arXiv:nucl-th/0302071 [nucl-th]].

\bibitem{STAR:2015mki}
L.~Adamczyk \textit{et al.} [STAR],
Phys. Rev. Lett. \textbf{115} (2015) no.22, 222301
[arXiv:1505.07812 [nucl-ex]].

\bibitem{Bernhard:2016tnd}
J.~E.~Bernhard, J.~S.~Moreland, S.~A.~Bass, J.~Liu and U.~Heinz,
Phys. Rev. C \textbf{94} (2016) no.2, 024907
[arXiv:1605.03954 [nucl-th]].

\bibitem{Nijs:2020ors}
G.~Nijs, W.~van der Schee, U.~G{\"u}rsoy and R.~Snellings,
Phys. Rev. Lett. \textbf{126} (2021) no.20, 202301
[arXiv:2010.15130 [nucl-th]].

\bibitem{Nijs:2020roc}
G.~Nijs, W.~van der Schee, U.~G{\"u}rsoy and R.~Snellings,
Phys. Rev. C \textbf{103} (2021) no.5, 054909
[arXiv:2010.15134 [nucl-th]].

\bibitem{JETSCAPE:2020mzn}
D.~Everett \textit{et al.} [JETSCAPE],
Phys. Rev. C \textbf{103} (2021) no.5, 054904
[arXiv:2011.01430 [hep-ph]].

\bibitem{Parkkila:2021yha}
J.~E.~Parkkila, A.~Onnerstad, S.~F.~Taghavi, C.~Mordasini, A.~Bilandzic, M.~Virta and D.~J.~Kim,
Phys. Lett. B \textbf{835} (2022), 137485
[arXiv:2111.08145 [hep-ph]].

\bibitem{Liyanage:2023nds}
D.~Liyanage, {\"O}.~S{\"u}rer, M.~Plumlee, S.~M.~Wild and U.~Heinz,
Phys. Rev. C \textbf{108} (2023) no.5, 054905
[arXiv:2302.14184 [nucl-th]].

\bibitem{Giacalone:2023cet}
G.~Giacalone, G.~Nijs and W.~van der Schee,
Phys. Rev. Lett. \textbf{131} (2023) no.20, 20
[arXiv:2305.00015 [nucl-th]].

\bibitem{Virta:2024avu}
M.~Virta, J.~Parkkila and D.~J.~Kim,
Phys. Rev. C \textbf{111} (2025) no.4, 044903
[arXiv:2411.01932 [hep-ph]].

\bibitem{Jaiswal:2025deb}
S.~Jaiswal,
Phys. Lett. B \textbf{874} (2026), 140243
[arXiv:2509.19759 [hep-ph]].

\bibitem{Eskola:1999fc}
K.~J.~Eskola, K.~Kajantie, P.~V.~Ruuskanen and K.~Tuominen,
Nucl. Phys. B \textbf{570} (2000), 379-389
[arXiv:hep-ph/9909456 [hep-ph]].

\bibitem{Eskola:2001bf}
K.~J.~Eskola, P.~V.~Ruuskanen, S.~S.~Rasanen and K.~Tuominen,
Nucl. Phys. A \textbf{696} (2001), 715-728
[arXiv:hep-ph/0104010 [hep-ph]].

\bibitem{Garcia-Montero:2025hys}
O.~Garcia-Montero and S.~Schlichting,
Eur. Phys. J. A \textbf{61} (2025) no.3, 54
[arXiv:2502.09721 [hep-ph]].

\bibitem{Lappi:2006fp}
T.~Lappi and L.~McLerran,
Nucl. Phys. A \textbf{772} (2006), 200-212
[arXiv:hep-ph/0602189 [hep-ph]].

\bibitem{Artru:1974hr}
X.~Artru and G.~Mennessier,
Nucl. Phys. B \textbf{70} (1974), 93-115

\bibitem{Andersson:1983ia}
B.~Andersson, G.~Gustafson, G.~Ingelman and T.~Sjostrand,
Phys. Rept. \textbf{97} (1983), 31-145

\bibitem{Lappi:2006hq}
T.~Lappi,
Phys. Lett. B \textbf{643} (2006), 11-16
[arXiv:hep-ph/0606207 [hep-ph]].

\bibitem{Berges:2020fwq}
J.~Berges, M.~P.~Heller, A.~Mazeliauskas and R.~Venugopalan,
Rev. Mod. Phys. \textbf{93} (2021) no.3, 035003
[arXiv:2005.12299 [hep-th]].

\bibitem{Jankowski:2020itt}
J.~Jankowski, S.~Kamata, M.~Martinez and M.~Spali{\'n}ski,
Phys. Rev. D \textbf{104} (2021) no.7, 074012
[arXiv:2012.02184 [nucl-th]].

\bibitem{Blaizot:2019scw}
J.~P.~Blaizot and L.~Yan,
Annals Phys. \textbf{412} (2020), 167993
[arXiv:1904.08677 [nucl-th]].

\bibitem{Borghini:2022iym}
N.~Borghini, M.~Borrell, N.~Feld, H.~Roch, S.~Schlichting and C.~Werthmann,
Phys. Rev. C \textbf{107} (2023) no.3, 034905
[arXiv:2209.01176 [hep-ph]].

\bibitem{Busza:2018rrf}
W.~Busza, K.~Rajagopal and W.~van der Schee,
Ann. Rev. Nucl. Part. Sci. \textbf{68} (2018), 339-376
[arXiv:1802.04801 [hep-ph]].

\bibitem{Giacalone:2019ldn}
G.~Giacalone, A.~Mazeliauskas and S.~Schlichting,
Phys. Rev. Lett. \textbf{123} (2019) no.26, 262301
[arXiv:1908.02866 [hep-ph]].

\bibitem{Miller:2007ri}
M.~L.~Miller, K.~Reygers, S.~J.~Sanders and P.~Steinberg,
Ann. Rev. Nucl. Part. Sci. \textbf{57} (2007), 205-243
[arXiv:nucl-ex/0701025 [nucl-ex]].

\bibitem{Loizides:2014vua}
C.~Loizides, J.~Nagle and P.~Steinberg,
SoftwareX \textbf{1-2} (2015), 13-18
[arXiv:1408.2549 [nucl-ex]].

\bibitem{Moreland:2018gsh}
J.~S.~Moreland, J.~E.~Bernhard and S.~A.~Bass,
Phys. Rev. C \textbf{101} (2020) no.2, 024911
[arXiv:1808.02106 [nucl-th]].

\bibitem{Kirchner:2025yuo}
A.~Kirchner and S.~A.~Bass,
[arXiv:2508.20390 [hep-ph]].

\bibitem{Nijs:2022rme}
G.~Nijs and W.~van der Schee,
Phys. Rev. Lett. \textbf{129} (2022) no.23, 232301
[arXiv:2206.13522 [nucl-th]].

\bibitem{Giacalone:2022hnz}
G.~Giacalone,
[arXiv:2208.06839 [nucl-th]].

\bibitem{Carzon:2021tif}
P.~Carzon, M.~D.~Sievert and J.~Noronha-Hostler,
Phys. Rev. C \textbf{105} (2022) no.1, 014913
[arXiv:2106.02525 [nucl-th]].

\bibitem{Bernhard:2019bmu}
J.~E.~Bernhard, J.~S.~Moreland and S.~A.~Bass,
Nature Phys. \textbf{15} (2019) no.11, 1113-1117

\bibitem{Hanus:2019fnc}
P.~Hanus, A.~Mazeliauskas and K.~Reygers,
Phys. Rev. C \textbf{100} (2019) no.6, 064903
[arXiv:1908.02792 [hep-ph]].

\bibitem{Bozek:2012fw}
P.~Bozek and W.~Broniowski,
Phys. Rev. C \textbf{85} (2012), 044910
[arXiv:1203.1810 [nucl-th]].

\bibitem{Voloshin:2007pc}
S.~A.~Voloshin, A.~M.~Poskanzer, A.~Tang and G.~Wang,
Phys. Lett. B \textbf{659} (2008), 537-541
[arXiv:0708.0800 [nucl-th]].

\bibitem{Bzdak:2012ab}
A.~Bzdak and V.~Koch,
Phys. Rev. C \textbf{86} (2012), 044904
[arXiv:1206.4286 [nucl-th]].

\bibitem{STAR:2013gus}
L.~Adamczyk \textit{et al.} [STAR],
Phys. Rev. Lett. \textbf{112} (2014), 032302
[arXiv:1309.5681 [nucl-ex]].

\bibitem{Rogly:2018kus}
R.~Rogly, G.~Giacalone and J.~Y.~Ollitrault,
Phys. Rev. C \textbf{99} (2019) no.3, 034902
[arXiv:1809.00648 [nucl-th]].

\bibitem{Braun-Munzinger:2023gsd}
P.~Braun-Munzinger, K.~Redlich, A.~Rustamov and J.~Stachel,
JHEP \textbf{08} (2024), 113
[arXiv:2312.15534 [nucl-th]].

\bibitem{Roubertie:2025qps}
E.~Roubertie, M.~Verdan, A.~Kirchner and J.~Y.~Ollitrault,
Phys. Rev. C \textbf{111} (2025) no.6, 064906
[arXiv:2503.17035 [nucl-th]].

\bibitem{Gardim:2019xjs}
F.~G.~Gardim, G.~Giacalone, M.~Luzum and J.~Y.~Ollitrault,
Nature Phys. \textbf{16} (2020) no.6, 615-619
[arXiv:1908.09728 [nucl-th]].

\bibitem{Gardim:2020sma}
F.~G.~Gardim, G.~Giacalone, M.~Luzum and J.~Y.~Ollitrault,
Nucl. Phys. A \textbf{1005} (2021), 121999
[arXiv:2002.07008 [nucl-th]].

\bibitem{Mu:2025gtr}
Y.~S.~Mu, J.~A.~Sun, L.~Yan and X.~G.~Huang,
Phys. Rev. Lett. \textbf{135} (2025) no.16, 162301
[arXiv:2501.02777 [nucl-th]].

\bibitem{Parida:2024ckk}
T.~Parida, R.~Samanta and J.~Y.~Ollitrault,
Phys. Lett. B \textbf{857} (2024), 138985
[arXiv:2407.17313 [nucl-th]].

\bibitem{SoaresRocha:2024drz}
G.~Soares Rocha, L.~Gavassino, M.~Singh and J.~F.~Paquet,
Phys. Rev. C \textbf{110} (2024) no.3, 034913
doi:10.1103/PhysRevC.110.034913
[arXiv:2405.10401 [hep-ph]].

\bibitem{ATLAS:2025nnt}
G.~Aad \textit{et al.} [ATLAS],
[arXiv:2509.05171 [nucl-ex]].

\bibitem{ALICE:2025luc}
I.~J.~Abualrob \textit{et al.} [ALICE],
[arXiv:2509.06428 [nucl-ex]].

\bibitem{CMS:2025tga}
A.~Hayrapetyan \textit{et al.} [CMS],
[arXiv:2510.02580 [nucl-ex]].

\bibitem{STAR:2025ivi}
 [STAR],
[arXiv:2510.19645 [nucl-ex]].

\bibitem{Jia:2022ozr}
J.~Jia, G.~Giacalone, B.~Bally, J.~D.~Brandenburg, U.~Heinz, S.~Huang, D.~Lee, Y.~J.~Lee, C.~Loizides and W.~Li, \textit{et al.}
Nucl. Sci. Tech. \textbf{35} (2024) no.12, 220
[arXiv:2209.11042 [nucl-ex]].

\bibitem{STAR:2024wgy}
M.~I.~Abdulhamid \textit{et al.} [STAR],
Nature \textbf{635} (2024) no.8037, 67-72
[arXiv:2401.06625 [nucl-ex]].

\bibitem{Giacalone:2025vxa}
G.~Giacalone, J.~Jia, V.~Som{\`a}, Y.~Zhou, A.~Afanasjev, M.~Alvioli, B.~Bally, F.~Capellino, J.~P.~Ebran and H.~Elfner, \textit{et al.}
[arXiv:2507.01454 [nucl-ex]].

\bibitem{Blaizot:2014wba}
J.~P.~Blaizot, W.~Broniowski and J.~Y.~Ollitrault,
Phys. Rev. C \textbf{90} (2014) no.3, 034906
[arXiv:1405.3274 [nucl-th]].

\bibitem{Luzum:2023gwy}
M.~Luzum, M.~Hippert and J.~Y.~Ollitrault,
Eur. Phys. J. A \textbf{59} (2023) no.5, 110
[arXiv:2302.14026 [nucl-th]].

\bibitem{PHOBOS:2006dbo}
B.~Alver \textit{et al.} [PHOBOS],
Phys. Rev. Lett. \textbf{98} (2007), 242302
[arXiv:nucl-ex/0610037 [nucl-ex]].

\bibitem{Alver:2010gr}
B.~Alver and G.~Roland,
Phys. Rev. C \textbf{81} (2010), 054905
[erratum: Phys. Rev. C \textbf{82} (2010), 039903]
[arXiv:1003.0194 [nucl-th]].



\end{thebibliography}
\end{document}